%% file: main.tex
\documentclass[conference]{./templates/IEEEtran}[conference]

\newcommand{\pdftitle}{Analyse or Transmit: Utilising Correlation at the Edge with Deep Reinforcement Learning}

\input{src/setup_and_packages}

\newcommand\copyrighttext{%
  \footnotesize \textcopyright 2021 IEEE. Personal use of this material is permitted.
  Permission from IEEE must be obtained for all other uses, in any current or future
  media, including reprinting/republishing this material for advertising or promotional
  purposes, creating new collective works, for resale or redistribution to servers or
  lists, or reuse of any copyrighted component of this work in other works.
  }
\newcommand\copyrightnotice{%
\begin{tikzpicture}[remember picture,overlay]
\node[anchor=south,yshift=10pt] at (current page.south) {\fbox{\parbox{\dimexpr\textwidth-\fboxsep-\fboxrule\relax}{\copyrighttext}}};
\end{tikzpicture}%
}

\begin{document}
\bstctlcite{IEEEexample:BSTcontrol}

\title{\pdftitle}

\author{\IEEEauthorblockN{Jernej Hribar\IEEEauthorrefmark{1},
Ryoichi Shinkuma\IEEEauthorrefmark{2}, George Iosifidis\IEEEauthorrefmark{3}, and
Ivana Dusparic\IEEEauthorrefmark{1}}
\IEEEauthorblockA{\IEEEauthorrefmark{1}CONNECT-Trinity College Dublin, Ireland, Email: \{jhribar, duspari\}@tcd.ie \\
\IEEEauthorrefmark{2}Faculty of Engineering, Shibaura Institute of Technology, Japan,   Email: shinkuma@shibaura-it.ac.jp\\
\IEEEauthorrefmark{3}Delft University of Technology, Netherlands,  Email: g.iosifidis@tudelft.nl\\
}}

\maketitle

\input{src/acronyms}

\begin{abstract}
Millions of sensors, cameras, meters, and other edge devices are deployed in networks to collect and analyse data. In many cases, such devices are powered only by \ac{eh} and have limited energy available to analyse acquired data. When edge infrastructure is available, a device has a choice: to perform analysis locally or offload the task to other resource-rich devices such as cloudlet servers. However, such a choice carries a price in terms of consumed energy and accuracy. On the one hand, transmitting raw data can result in a higher energy cost in comparison to the required energy to process data locally. On the other hand, performing data analytics on servers can improve the task's accuracy. Additionally, due to the correlation between information sent by multiple devices, accuracy might not be affected if some edge devices decide to neither process nor send data and preserve energy instead. For such a scenario, we propose a \ac{drl} based solution capable of learning and adapting the policy to the time-varying energy arrival due to \ac{eh} patterns. We leverage two datasets, one to model energy an \ac{eh} device can collect and the other to model the correlation between cameras. Furthermore, we compare the proposed solution performance to three baseline policies. Our results show that we can increase accuracy by $15\%$ in comparison to conventional approaches while preventing outages.
\end{abstract}

\copyrightnotice

\acresetall


\section{Introduction}
\label{sec:intro}
\input{src/sections/1_introduction}

\section{Related Work}
\label{sec:related}
\input{src/sections/2_related}
\section{System Model and Problem Statement}
\label{sec:system_model}
\input{src/sections/3_system_model}


\section{DRL-based Solution}
\label{sec:solution}
\input{src/sections/5_solution}

\section{Validation and Results}
\label{sec:validation}
\input{src/sections/6_validation}
\section{Conclusion}
\label{sec:conclusion}
\input{src/sections/7_conclusion}

\section*{Acknowledgements}

This work was funded in part by the European Regional Development Fund through the SFI Research Centres Programme under Grant No. 13/RC/2077\_P2 SFI CONNECT, the SFI-NSFC Partnership Programme Grant No. 17/NSFC/5224, and European Union’s H2020 research and innovation programme under grant agreement DAEMON 101017109.
This work was also supported in part by JST PRESTO Grant No. JPMJPR1854, JSPS KAKENHI Grant No. JP21H03427, and JSPS International Research Fellow Grant No. PE20723. 

\balance

\bibliographystyle{./templates/IEEEtran}
\bibliography{IEEEabrv,bibliography}

\end{document}

%% file: src/setup_and_packages.tex
\usepackage[nolist,nohyperlinks]{acronym}
\usepackage{amsmath}
\usepackage{amssymb} 
\usepackage{breqn}
\usepackage{balance}
\usepackage{cite}
\usepackage{framed}
\usepackage{siunitx}
\usepackage{soul}
\usepackage{multirow}
\usepackage{etoolbox}
\usepackage{standalone}
\usepackage{textcomp}
\usepackage{multirow,booktabs}
\usepackage{amsmath}
\usepackage{amssymb}
\usepackage{bm}
\usepackage{mathtools} 
\usepackage{optidef} 

\usepackage{xcolor, colortbl, collcell}
\colorlet{activecolour}{black}

\usepackage{graphicx}
\usepackage{subcaption}
\usepackage[font=small]{caption}

\graphicspath{{./images/}}

\usepackage{tikz}
\usepackage{pgfplots}
\usepgfplotslibrary{groupplots}

\usepackage{url}
\usepackage[bookmarks=false]{hyperref}
\hypersetup{
  backref,
  hidelinks,
  colorlinks=false,
  breaklinks=true,
  bookmarksopen=false,
  pdftitle=\pdftitle,
  pdfauthor={Author}
}

\newtoggle{authorcomment}
\toggletrue{authorcomment}

\definecolor{LightCyan}{rgb}{0.88,1,1}
\definecolor{Gray}{gray}{0.85}

\usetikzlibrary{arrows,decorations.pathmorphing,backgrounds,positioning,fit,petri}
\usetikzlibrary{shapes.misc}
\tikzset{cross/.style={cross out, draw=black, minimum size=2*(#1-\pgflinewidth), inner sep=0pt, outer sep=0pt},
	cross/.default={1pt}}
\usetikzlibrary{patterns}

%% file: src/acronyms.tex
\begin{acronym}[MACHU]
  \acro{iot}[IoT]{Internet of Things}
  \acro{cr}[CR]{Cognitive Radio}
  \acro{ofdm}[OFDM]{orthogonal frequency-division multiplexing}
  \acro{ofdma}[OFDMA]{orthogonal frequency-division multiple access}
  \acro{scfdma}[SC-FDMA]{single carrier frequency division multiple access}
  \acro{rbi}[RBI]{ Research Brazil Ireland}
  \acro{rfic}[RFIC]{radio frequency integrated circuit}
  \acro{sdr}[SDR]{Software Defined Radio}
  \acro{sdn}[SDN]{Software Defined Networking}
  \acro{su}[SU]{Secondary User}
  \acro{ra}[RA]{Resource Allocation}
  \acro{qos}[QoS]{quality of service}
  \acro{usrp}[USRP]{Universal Software Radio Peripheral}
  \acro{mno}[MNO]{Mobile Network Operator}
  \acro{mnos}[MNOs]{Mobile Network Operators}
  \acro{gsm}[GSM]{Global System for Mobile communications}
  \acro{tdma}[TDMA]{Time-Division Multiple Access}
  \acro{fdma}[FDMA]{Frequency-Division Multiple Access}
  \acro{gprs}[GPRS]{General Packet Radio Service}
  \acro{msc}[MSC]{Mobile Switching Centre}
  \acro{bsc}[BSC]{Base Station Controller}
  \acro{umts}[UMTS]{universal mobile telecommunications system}
  \acro{Wcdma}[WCDMA]{Wide-band code division multiple access}
  \acro{wcdma}[WCDMA]{wide-band code division multiple access}
  \acro{cdma}[CDMA]{code division multiple access}
  \acro{lte}[LTE]{Long Term Evolution}
  \acro{papr}[PAPR]{peak-to-average power rating}
  \acro{hn}[HetNet]{heterogeneous networks}
  \acro{phy}[PHY]{physical layer}
  \acro{mac}[MAC]{medium access control}
  \acro{amc}[AMC]{adaptive modulation and coding}
  \acro{mimo}[MIMO]{multiple input multiple output}
  \acro{rats}[RATs]{radio access technologies}
  \acro{vni}[VNI]{visual networking index}
  \acro{rbs}[RB]{resource blocks}
  \acro{rb}[RB]{resource block}
  \acro{ue}[UE]{user equipment}
  \acro{cqi}[CQI]{Channel Quality Indicator}
  \acro{hd}[HD]{half-duplex}
  \acro{fd}[FD]{full-duplex}
  \acro{sic}[SIC]{self-interference cancellation}
  \acro{si}[SI]{self-interference}
  \acro{bs}[BS]{base station}
  \acro{fbmc}[FBMC]{Filter Bank Multi-Carrier}
  \acro{ufmc}[UFMC]{Universal Filtered Multi-Carrier}
  \acro{scm}[SCM]{Single Carrier Modulation}
  \acro{isi}[ISI]{inter-symbol interference}
  \acro{ftn}[FTN]{Faster-Than-Nyquist}
  \acro{m2m}[M2M]{machine-to-machine}
  \acro{mtc}[MTC]{machine type communication}
  \acro{mmw}[mmWave]{millimeter wave}
  \acro{bf}[BF]{beamforming}
  \acro{los}[LOS]{line-of-sight}
  \acro{nlos}[NLOS]{non line-of-sight}
  \acro{capex}[CAPEX]{capital expenditure}
  \acro{opex}[OPEX]{operational expenditure}
  \acro{ict}[ICT]{information and communications technology}
  \acro{sp}[SP]{service providers}
  \acro{inp}[InP]{infrastructure providers}
  \acro{mvnp}[MVNP]{mobile virtual network provider}
  \acro{mvno}[MVNO]{mobile virtual network operator}
  \acro{nfv}[NFV]{network function virtualization}
  \acro{vnfs}[VNF]{virtual network functions}
  \acro{cran}[C-RAN]{Cloud Radio Access Network}
  \acro{bbu}[BBU]{baseband unit}
  \acro{bbus}[BBU]{baseband units}
  \acro{rrh}[RRH]{remote radio head}
  \acro{rrhs}[RRH]{Remote radio heads} 
  \acro{sfv}[SFV]{sensor function virtualization}
  \acro{wsn}[WSN]{wireless sensor networks} 
  \acro{bio}[BIO]{Bristol is open}
  \acro{vitro}[VITRO]{Virtualized dIstributed plaTfoRms of smart Objects}
  \acro{os}[OS]{operating system}
  \acro{www}[WWW]{world wide web}
  \acro{iotvn}[IoT-VN]{IoT virtual network}
  \acro{mems}[MEMS]{micro electro mechanical system}
  \acro{mec}[MEC]{Mobile edge computing}
  \acro{coap}[CoAP]{Constrained Application Protocol}
  \acro{vsn}[VSN]{Virtual sensor network}
  \acro{rest}[REST]{REpresentational State Transfer}
  \acro{aoi}[AoI]{Age of Information}
  \acro{lora}[LoRa\texttrademark]{Long Range}
  \acro{iot}[IoT]{Internet of Things}
  \acro{snr}[SNR]{Signal-to-Noise Ratio}
  \acro{cps}[CPS]{Cyber-Physical System}
  \acro{uav}[UAV]{Unmanned Aerial Vehicle}
  \acro{rfid}[RFID]{Radio-frequency identification}
  \acro{lpwan}[LPWAN]{Low-Power Wide-Area Network}
  \acro{lgfs}[LGFS]{Last Generated First Served}
  \acro{wsn}[WSN]{wireless sensor network} 
  \acro{lmmse}[LMMSE]{Linear Minimum Mean Square Error}
  \acro{rl}[RL]{Reinforcement Learning}
  \acro{nb-iot}[NB-IoT]{Narrowband IoT}
  \acro{lorawan}[LoRaWAN]{Long Range Wide Area Network}
  \acro{mdp}[MDP]{Markov Decision Process}
  \acro{ann}[ANN]{Artificial Neural Network}
  \acro{dqn}[DQN]{Deep Q-Network}
  \acro{mse}[MSE]{Mean Square Error}
  \acro{ml}[ML]{Machine Learning}
  \acro{cpu}[CPU]{Central Processing Unit}
  \acro{ddpg}[DDPG]{Deep Deterministic Policy Gradient}
  \acro{ai}[AI]{Artificial Intelligence}
  \acro{gp}[GP]{Gaussian Processes}
  \acro{drl}[DRL]{Deep Reinforcement Learning}
  \acro{mmse}[MMSE]{Minimum Mean Square Error}
  \acro{fnn}[FNN]{Feedforward Neural Network}
  \acro{eh}[EH]{Energy Harvesting}
  \acro{wpt}[WPT]{Wireless Power Transfer}
  \acro{dl}[DL]{Deep Learning}
  \acro{yolo}[YOLO]{You Only Look Once}
  \acro{mec}[MEC]{Mobile Edge Computing}
\end{acronym}

%% file: src/sections/1_introduction.tex

In the last few years, we have witnessed unprecedented progress in \ac{ml} methods that paved the way for many new analytic services where embedded devices collect and process data to extract information of interest~\cite{siow2018analytics}. Coupled with the rise of available processing power at the network edge through cloudlet servers~\cite{chen2019deep}, applications that require data analysis such as cameras counting objects or sensors detecting anomalies have become the new norm. In such a scenario, devices are confronted with a decision: to analyse collected data locally (and transmit only the extracted information of interest) or transmit obtained data to a fusion centre for processing. Such a decision carries a cost in terms of device's energy consumption and applications' performance, e.g., the accuracy of detected anomaly. 

Providing energy necessary to power edge devices is becoming increasingly more challenging~\cite{green_survey}. While connecting devices to the power grid is the most reliable way of providing energy, it also results in high deployment costs. On the one hand, \ac{eh} is proving a viable alternative as it enables the device to collect energy by tapping into the various ambient sources such as wind, solar radiation, vibrations, etc.\cite{bi2015wireless}. On the other hand, a device with an \ac{eh} has to be aware of its energy usage while ensuring that the application performs as required. Additionally, the collected energy by an \ac{eh} varies over time. For example, during the day, a device with a solar panel can gather plenty of energy, while during the night, collected energy will be negligible. 

In large sensor networks, the information obtained from individual devices is often correlated. For example, devices might be monitoring overlapping areas and  multiple devices can detect the same anomaly or the observing events are correlated. Therefore, redundancy creates an opportunity to reduce the operational costs of deployments. In other words, an \ac{eh}-powered device can take advantage of the correlation in order to improve its energy efficiency. For example, a device can choose not to take any actions and preserve energy instead, as another device will obtain the required information. However, to enable such an approach, the system has to learn how to take advantage of correlated information.

Existing work mostly focuses on a single aspect of this problem. For example,\cite{min2019learning, balasubramanian2018unified, liu2016energy, xu2017online} consider offloading computational task from an \ac{eh}-powered device to a server, but without considering the accuracy of the task. Work in \cite{lyu2018selective, ran2017delivering, optimizing_data_analytics}, on the other hand, considers data-analytics scenarios, but does not consider \ac{eh}-powered devices. In addition, neither of the above approaches consider taking advantage of correlation. 

To address this gap, in this paper we investigate the impact of correlation in the outcome of the data-analytic task on the energy efficiency of \ac{eh}-powered devices at the network edge. To examine the characteristics of such a system, we focus on a case study of a set of cameras powered by \ac{eh} counting traffic in an intersection. Cameras can choose to perform object detection or transmit a raw image to be processed on a nearby cloudlet server. Additionally, due to correlation, one of the cameras can decide not to transmit or perform object detection because the system will be able to count vehicles accurately using an image from the other camera. In such a scenario, object detection can be performed by only a subset of all available cameras while it is still possible to detect all objects. The system's goal is to count observed objects accurately, which we measure through a standard \ac{ml} metric recall while being limited by the time-varying energy of cameras with \ac{eh}. We identified \ac{drl} as the most suitable approach to design an autonomous solution capable of adapting to varying energy of a device with \ac{eh} while taking advantage of correlated information. The benefits of learning approach in such environments was demonstrated by many other \ac{drl}-based solutions~\cite{luong2018applications, aoudia2018rlman, chu2018reinforcement, li2019deep}. The contributions in this paper can be summarised as follows:

\begin{itemize}

\item We propose and formulate the problem of joint energy and accuracy optimization in \ac{eh} edge networks with correlated sources of information.

\item We propose a novel \ac{drl}-based algorithm that guides the decision for  \ac{eh}-powered sensors based on available energy, task accuracy, and correlation.

\item We enrich the \ac{drl} solution with a \ac{gp}-based module for predicting on the fly the devices' energy availability.

\item We evaluate the performance of the two variants of the \ac{drl}-based solutions (with and without \ac{gp}) using two established datasets\cite{video_dataset, harvesting_dataset} to simulate the environment as accurately as possible and compare our solution to three baseline policies.

\end{itemize}

%% file: src/sections/2_related.tex

Our work is related to studies that proposed task offloading schemes for edge devices powered by an \ac{eh}~\cite{min2019learning, balasubramanian2018unified, liu2016energy, xu2017online}.
For example, the work in \cite{min2019learning} leverages \ac{drl} to decide which edge device will offload the task so as to minimize their energy consumption and task latency. The authors in~\cite{balasubramanian2018unified} propose an architecture and a threshold policy to achieve energy-aware edge task offloading from \ac{eh} sensors. Similarly in \cite{liu2016energy}, the authors rely on online Lyapunov based task offloading algorithm to investigate the trade-off between energy consumption and execution delay. And~\cite{xu2017online} considered online learning to decide on how much power should a \ac{mec}, powered by an \ac{eh}, allocate to each task. However, the above solutions do not optimise the accuracy of the data-analytics task, nor aim to design a system that will ensure that \ac{eh} powered devices will avoid depleting all available energy as is the objective of our work.

Only a handful of papers examine the optimisation of task offloading schemes to improve the accuracy of data-analytics~\cite{lyu2018selective, ran2017delivering, optimizing_data_analytics}. In \cite{lyu2018selective} the authors use facial recognition as an example to validate the energy effectiveness of their proposed task offloading scheme. However, in their system \ac{eh} is not considered. Similarly, a trade-off between object detection accuracy and latency was explored in~\cite{ran2017delivering} for battery-powered mobile devices. The focus of the latter was to demonstrate that task offloading can improve frame rate and accuracy. An online learning approach was proposed in \cite{optimizing_data_analytics}, which improves the accuracy of the data-analytics task while reducing the energy consumption of an edge device. In contrast, we explore a more advanced system in which devices are powered only by an \ac{eh}. In our system, when deciding the system must also consider the energy cameras will collect in future and correlation between cameras to achieve optimal performance.

Recently, \acf{rl} emerged as a very effective tool to resolve a plethora of problems related to the management of \ac{eh} devices due to its ability to adapt to a dynamic environment~\cite{aoudia2018rlman, chu2018reinforcement, li2019deep}. For example, in \cite{aoudia2018rlman} the authors proposed a \ac{rl}-based power management capable of maximising the quality of service while considering available energy and energy cost of transmission. In \cite{chu2018reinforcement} the authors employed \ac{drl} to resolve an access problem for \ac{eh} devices. A balance between transmission power and modulation level to increase throughput was considered in \cite{li2019deep}. In the papers reviewed above, \ac{drl} was selected because of its adaptability to the time-varying nature of energy arrival on \ac{eh}-powered devices, leading us to consider it as a suitable approach for the problem our paper is addressing as well.

%% file: src/sections/3_system_model.tex

\begin{figure}
	\centering
	\includegraphics[width=3.0in]{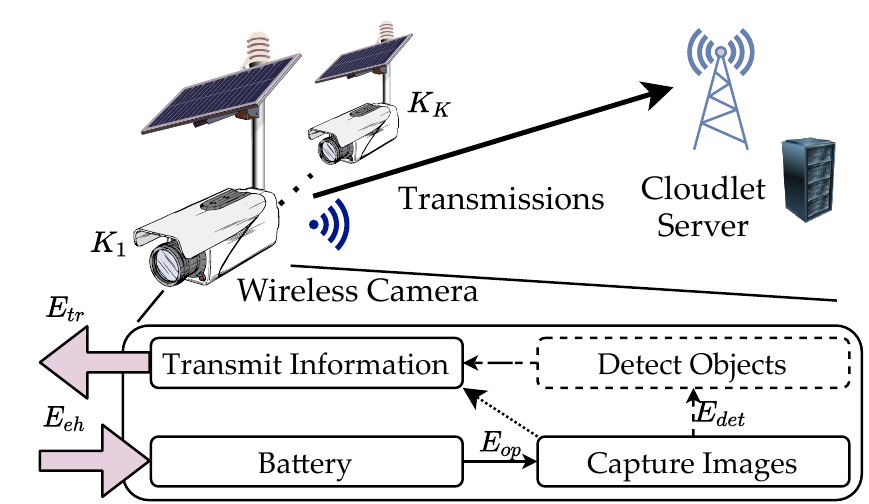}
	\caption{Illustration of the system model and the main wireless camera components.}
	\label{fig:system_model}
	\vspace{-15pt}
\end{figure}

We consider a network consisting of $K$ wireless cameras powered by an \ac{eh} and a single cloudlet server as illustrated in Fig. \ref{fig:system_model}.
The cameras are embedded devices capable of processing frames, i.e., images, using dedicated hardware, and transmitting gained information to the cloudlet server for collection. Alternatively, a camera can transmit captured frames and allow the cloudlet server to process them. We assume that time is slotted $t \in \mathcal{T} = \{0, 1, \hdots, T -1 \} $ and at each time slot each cameras captures one image. The system updates its decision every $\tau$ time-steps as due to relatively high cameras frame rate\footnote{Typically traffic video cameras capture 10 frames per second\cite{naphade2019ai}} $\lambda$ deciding every time-step would be impractical.

\subsection{Cameras' Energy Parameters}

In each time-step, the $k$-th camera receives harvested energy $E_{eh}^{(k)} (t)$ proportional to the current $I_{eh}^{(k)}(t)$ and the voltage $U_{eh}^{(k)} (t)$ from the \acf{eh} unit. To store the captured energy, each cameras is quipped with a battery that has a maximal capacity $E_{max}$. Meaning, camera's available energy $E_{av}^{(k)}(t)$ is limited to an interval $E_{av}^{(k)} \in [0, E_{max}]$. Note that if camera's energy is zero, i.e, $E_{av}^{(k)}(t) \leq 0$,  camera turns-off and no images nor detected objects are available to the system.
We denote the downtime of $k$-th camera as $T_{down}^{(k)}$.

Operating the camera consumes $E_{op}(t)$ of energy at each time step. This energy is required to support the camera's essential operation, such as capturing images. We assume that the system has no control over $E_{op}(t)$ energy consumption regardless of how the camera operates. Therefore, even if the system decides that no images are required from the camera, the camera will still consume $E_{op}(t)$. The $E_{op}(t)$ depends mainly on the hardware used. In the validation section, we model $E_{op}(t)$ according to the consumption of a Raspberry Pi with a camera.

Whenever the camera decides to transit the image, the energy consumption is relative to the raw image size in bits we denote as $f_{raw}^{(k)}(t)$ multiplied by the energy cost of transmitting a bit of information $E_{tr}(t)$. To detect the object locally, we assume that the camera consumes $E_{det}(t)$ of energy. Such energy is required to process the image. Furthermore, the camera will also consume a small amount of energy to transmit obtained information, e.g., the number of detected objects. We denote the size of such data packet with $f_{proc}^{(k)}(t)$. However, the latter data packet is much smaller in size than when camera decides to transmit raw image, i.e, $f_{raw}^{(k)}(t) > f_{proc}^{(k)}(t)$.

\subsection{Object Detection, Correlation, and Recall}

We define $\mathcal{C}(t)$ as the set of all objects present at a particular time-step $t$ in the scenery of interest, e.g., the intersection. A camera can not always detect all objects due to practical limitations, thus we define a set of all possible objects a camera can detects as $\mathcal{C}_k(t)$ which is a subset of $\mathcal{C}(t)$, i.e., $\mathcal{C}_k(t) \subseteq \mathcal{C}(t)$.
We assume that a device can process captured images using dedicated hardware capable of efficiently processing image. Edge TPU, NVIDIA Xavier, and NovuTenso are examples of such dedicated hardware units capable of ensuring that the energy cost of processing on edge device is lower than the energy cost of transmitting captured images\cite{hui2019early}. 
To detect objects of interest, the system can choose to employ object detector $J_{yolo}$ located on the cloudlet or $J_{tiny}$ on the camera. We use \ac{yolo} version 3\cite{yolo_reference} for object detection on the cloudled and tiny-yolo version 4 to perform object detection on the camera. 
Using one of the available detectors, the system can then obtain a set of objects detected by the individual cameras as follows:

\begin{equation}\label{eq:detection}
J_{yolo}:\mathcal{C}_k(t) \to  \mathcal{K}_k(t) \phantom{a} or   \phantom{a} J_{tiny}:\mathcal{C}_k(t) \to  \mathcal{K}_k(t),
\end{equation}

\noindent where $\mathcal{K}_k(t) $ is the set of object that are detected in the image of camera $k$ at time step $t$. Note, that if camera is in the stand-by mode or has no available energy, the resulting set is empty. Finally, we can define the set of all the objects the system has detected as $ \mathcal{K}(t) = \mathcal{K}_1(t)\cup\mathcal{K}_2(t)\hdots\cup \mathcal{K}_K(t)$. We also define the set of correlated objects as $\mathcal{C}_\rho(t) = \mathcal{C}_1(t)\cap\hdots\cap \mathcal{C}_K(t)$, i.e., the set of objects that all cameras can detect. 

To measure the performance of the system we adopt recall, a standard metric in \ac{ml} applications used to measure how well can a system detect relevant objects. In our case, the relevant objects are in $\mathcal{C}(t)$. Therefore, we define recall as:

\begin{equation}\label{eq:recall}
\phi (t) = \frac{|\mathcal{K}|}{|\mathcal{C}(t)|}.
\end{equation}

\noindent Recall is limited to an interval $\phi(t) \in [0,1]$. The higher the recall value the more objects the system detected. 

\subsection{Problem Formulation} 

The main objective of the system is to find a policy $\pi$, that decides where or if should camera's image be processed, that will maximise recall $\phi$. We formulate the problem as:

\begin{equation}
\begin{aligned}
\max_{\pi}  \quad &  \overline{\phi}(t)\\
\textrm{s.t.}  \quad & E_{av}^{(k)} (t)  \geq 0, \forall \: k=1,\ldots,K \\
& T_{down}^{(k)} \leq 0, \forall \: k=1,\ldots,K \\
  & t \in \mathcal{T} = \{0, 1, \hdots, T -1 \}   \\
\end{aligned}
\label{eq:problem_formulation}
\end{equation}

\noindent The main constraint that policy faces is limited available energy which varies over time. Simultaneously, the system minimise cameras' outages, i.e., $T_{down}=0$.

%% file: src/sections/5_solution.tex

\begin{figure}
	\centering
	\includegraphics[width=2.8in]{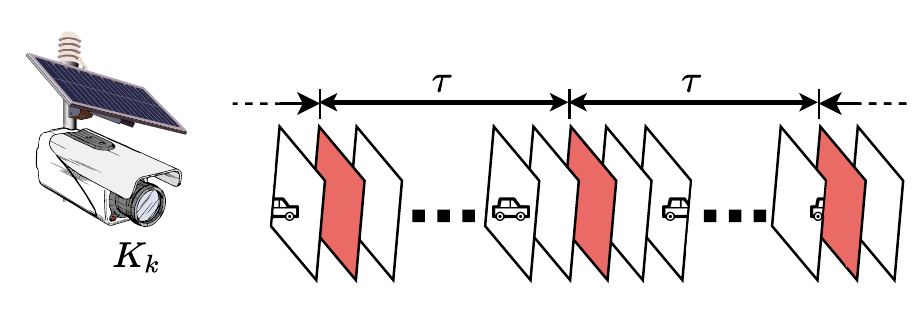}
	\caption{At each $\tau$ decision interval, all cameras will transmit the raw image to the cloudlet server.}
	\label{fig:guard_interval}
	\vspace{-15pt}
\end{figure} 

Finding a policy capable of maximising accuracy and preserve cameras' energy in a time-varying environment is a non-trivial task. Therefore, we base our solution on \ac{drl} , which recently emerged as an ideal tool for optimising the system's performance in such an environment~\cite{luong2018applications} due to its adaptability.
Furthermore, the system also has to overcome a practical challenge. In deployments, $\mathcal{C}(k)$ is unknown to the system. To tackle the issue, we propose the use of the guard interval, i.e., at every decision epoch $\tau$ all cameras transmit a raw image as illustrated in Fig. \ref{fig:guard_interval}. The system then utilises $J_{yolo}$ to determine the $\mathcal{K}(\tau)$, which for the proposed solution represents the real state of the system on which all decisions are based. In our preliminary studies in which we analysed the video dataset\cite{video_dataset} we discovered that using the $J_{yolo}$ on images from all available cameras at time $t$ results in $\overline{\phi}(t) =0.99$. We then use the information extracted from guard images to determine the states, actions, and rewards, i.e., a tuple in $\langle \mathcal{S}, \mathcal{A}, R \rangle$.

\subsection{States, Actions, and Reward}

The state $\mathbf{s}(t) \in \mathcal{S} \subseteq R^{K+2}$ comprises the residual energy in each device, the guard recall ($\phi_g(t)$) and the number of the detected objects, i.e., $|\mathcal{K}(\tau)|$. We determine the $\phi_g(t)$ by comparing the number of objects we counted one time-step before the guard interval as follows:

\begin{equation}\label{eq:recall_guard}
\phi_g (\tau) = \frac{|\mathcal{K}(\tau-1)|}{|\mathcal{K}(\tau)|}.
\end{equation}

\noindent Such a state space, even with a low granularity of discretising the inputs, rises extremely quickly to a thousands of states. For example, even using only a hundred states for the energy level, and relatively low granularity of $0.025$ for the recall state, for $K=4$, the number of possible states is in the millions. This is a rationale for designing a solution based on deep rather than tabular \ac{rl}.

Each camera has three different modes of operation: transmit raw image to the cloudlet for processing, use local object detector, or enter stand-by mode. Each action $\mathbf{a}$ in the set of available actions $\mathcal{A}$ represent a vector with dimension $K$, consisting of one operation mode per sensor.  

The reward consists of two parts. The first part is based on the observed recall value in a time-step before the decision time-step. To distinguish between acceptable rates and unsatisfactory ones, this reward is expressed as $r_{\phi}= 10(\phi_g(\tau)-0.5)$, meaning that any recall less than $0.5$ results in a negative reward, while only those higher than $0.5$ result in a positive one. To further accelerate learning, the resulting reward is multiplied by $10$. The second part of the reward depends on the cameras' energy; if any of the cameras have less than 15\% the agent receives a high negative reward.

\subsection{Estimating Energy and Implementation}

To estimate the energy the camera collects through \ac{eh} we use \acf{gp}~\cite{roberts2013gaussian}. Namely we selected a standard periodic kernel in combination with a white noise kernel. We fit \ac{gp} model using measurements of $I_{eh}^{(k)}(t)$ the system can easily access. 
Additionally, fitting the \ac{gp} model for estimation can be carried out only once per day thus adding minimal overhead in terms of required processing power. However, the dimension of the input state vector $\mathbf{s}(t)$ increases to $2K+2$, i.e., $\mathcal{S} \subseteq R^{2K+2}$. The new state space has to also encompass the information regarding the estimated energy.
In the next section, we demonstrate that using \ac{gp} to estimate energy can greatly aid the system in reducing the $T_{down}$ while also positively impacting recall.

\begin{table}[ht]
	\centering
	\caption{DQN Hyperparameters}
	\label{table_hyper_param}
	\input{tables/DQN_parameters}
	\vspace{-8pt}
\end{table}

We implemented the learning agent in the cloudlet server where enough computational power is available to support processing required to support \ac{dqn} and \ac{gp}.
Our \ac{dqn} implementation uses an \ac{ann} that consist of four hidden layers. The first and the last hidden layer have four neurons, while the middle two layers have eight neurons. We use the ReLU activation function for all layers except for the output layer, activated with a linear function. To prevent over-fitting, we apply a ten percent dropout between the layers.  We list the rest of the hyperparameters in Table \ref{table_hyper_param}.

%% file: tables/DQN_parameters.tex
\begin{tabular}{llll}
		\toprule
		\begin{tabular}[c]{@{}c@{}} Hyperparameter\end{tabular} & 
		\begin{tabular}[c]{@{}c@{}} Value \end{tabular} &
		\begin{tabular}[c]{@{}c@{}} Hyperparameter\end{tabular} & 
		\begin{tabular}[c]{@{}c@{}} Value \end{tabular} \\
		\arrayrulecolor{black}\hline 
		\midrule 
		
		\begin{tabular}[c]{@{}c@{}}  Batch size   \end{tabular} & 
		\begin{tabular}[c]{@{}c@{}}   $128$ \end{tabular} &
		
		\begin{tabular}[c]{@{}c@{}}  Memory size  \end{tabular} & 
		\begin{tabular}[c]{@{}c@{}}   $2*10^5$\end{tabular} \\

		\begin{tabular}[c]{@{}c@{}}  Optimizer  \end{tabular} & 
		\begin{tabular}[c]{@{}c@{}}   Adam \end{tabular} &
		\begin{tabular}[c]{@{}c@{}}  Loss Function  \end{tabular} & 
		\begin{tabular}[c]{@{}c@{}}   MSE \end{tabular} \\

		\begin{tabular}[c]{@{}c@{}}  Target ANN soft update $\tau$  \end{tabular} & 
		\begin{tabular}[c]{@{}c@{}}   $10^{-3}$\end{tabular} &
		
				\begin{tabular}[c]{@{}c@{}}   \end{tabular} & 
		\begin{tabular}[c]{@{}c@{}}  \end{tabular} \\
		
		\bottomrule   
\end{tabular}

%% file: src/sections/6_validation.tex

In this section, we validate the performance of the proposed solution using real observations from two datasets. We test two variations of our \ac{drl}-based approach, one with estimated arrival energy we obtain using \ac{gp} and one without such information. For comparison, we designed three heuristic approaches: greedy, threshold, and alternating. We evaluate the performance in two scenarios: static and dynamic varying, cost of transmitting a bit. We simulate the performance over fifteen days, but the reported results are based on the average of the last twelve days. Our experiments show that the \ac{drl}-based solution will start outperforming a random approach within one day and then requires another two to three days to converge to the best policy the agent can learn. Additionally, to minimise the impact of randomness on results, we report the average of five iterations for each tested approach.

\subsection{Use of Real-data and Heuristic Approaches}

The Ko-PER intersection dataset~\cite{video_dataset} comprises of mono-chrome camera images and raw laser scanner measurements for an intersection. The data sequence is six and half minutes long or 9670 frames obtained from two different viewpoints, i.e., cameras. Consequently, in this evaluation, we had to limit the number of cameras to two, i.e., $K=2$. Furthermore, we extrapolate the data by re-sampling. The Ko-PER images are obtained with 25 frames per second. However, in our simulation, we set the $\lambda$ to ten as is the standard for traffic camera\cite{naphade2019ai}.
To avoid correlation between the samples, every minute we randomly switch the starting point in the sequence from which our simulation receives frames. Fig. 3 shows an example of how our system counts vehicles. To obtain $\mathcal{C}(t)$, we manually counted vehicles inside the intersection. Note that only the white vehicle in the bottom left, as seen from camera one, i.e., Fig. 3(a), has entered the intersection. The same applies to vehicles on top of the image from camera two, i.e., Fig. 3(b), those vehicles have already left or have not entered the intersection yet. Meaning that $\mathcal{C}(t)=3$. The resulting $\mathcal{K}_1(t) $ and $\mathcal{K}_2(t) $ would be two and three, respectively. 


\begin{figure}
    \begin{subfigure}[b]{0.24\textwidth}
        \includegraphics[width=\textwidth]{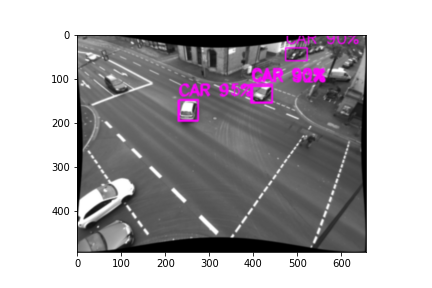}
        \caption[]{{\small Camera one.}}
    \end{subfigure}
   \begin{subfigure}[b]{0.24\textwidth}  
        \includegraphics[width=\textwidth]{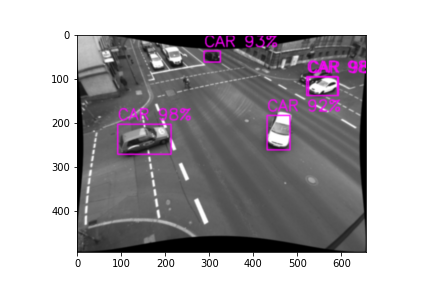}
        \caption[]{{\small Camera two.}}
    \end{subfigure}
    \caption{Example of images obtained from Ko-PER dataset at the same time-step from both cameras and objects detected by $J_{yolo}$.}
    \vspace{-15pt}
\end{figure}

We obtain the $I_{eh}^{(k)}(t)$ values from the energy-harvesting testbed installed on UCLM, Ciudad Real (Spain)\cite{harvesting_dataset}. We use measurements obtained between the $16^{th}$ and $30^{th}$ of August 2018 as input to our simulation, which enables us to test the performance of the proposed scheme over $15$ days. The nominal power of used solar panels is $2W$, and in practice, it is expected that multiple panels are used to power a wireless camera. We assume that each of the cameras is equipped with four panels with $80\%$ efficiency, i.e., $\eta_{eh}=0.8$. We list the rest of the simulation parameters in Table \ref{table_sim_param} which we modelled according to existing measurements. For example, $E_{det}$ we determined according to measurements in~\cite{hui2019early}. Furthermore, we also use real measurement to fit the \ac{gp} model to estimate collected energy. After performing empirical studies considering the intervals for model fitting and estimated energy, we determined that using the past seven days of data and estimating the amount of energy for the next six hours is most effective. The resulting approach can very accurately estimate the amount of energy a camera will collect in the future\footnote{A python notebook with detailed analysis can be found here: \href{https://github.com/hribarjernej89/estimating-eh-energy-with-GP}{github.com/hribarjernej89/estimating-eh-energy-with-GP}.}.

\begin{table}[ht]
	\centering
	\caption{Simulation Parameters}
	\label{table_sim_param}
	\input{tables/simulation_parameters}
	\vspace{-10pt}
\end{table}

We compare the performance of our proposed solution to three heuristic approaches:
\begin{enumerate}
	\item \textbf{Greedy}: When cameras adopt greedy policy they will always transmit the image, provided sufficient amount of energy is available. 
	\item \textbf{Threshold}:  The camera will transmit the image if it has more than half of the overall energy available, i.e., $E_{av}(t)^{k} E_{max}$. Otherwise, the camera will perform object detection locally and then transmit results.
	\item \textbf{Alternating}: Cameras iterate over $\mathcal{A}$ set of actions. At each decision epoch $\tau$ the operating mode of cameras changes.
\end{enumerate}

\begin{figure}
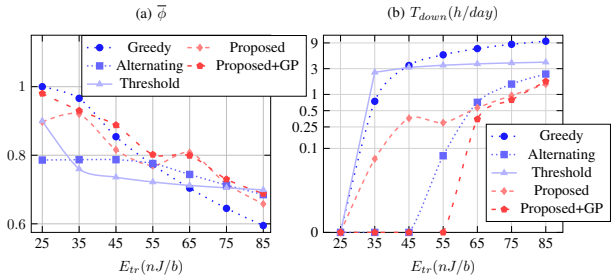

	\centering
	\includestandalone[width=0.45\textwidth]{tikz_figures/static_cost_of_transmitting}
	\caption{Recall and downtime depending on the energy cost of transmission for different approaches.}
	\label{fig:static}
	\vspace{-15pt}
\end{figure}

\subsection{Static Cost of Transmission}

Fig. \ref{fig:static}(a) shows the average recall value the system achieves depending on the energy cost of transmission. Both of our proposed solutions have a very good performance in terms of achieved recall. When the cost of transmission increases, the performance of all approaches performance decreases, as expected. The recall is lower when the cost of transmission is high because the proposed solution must more often select an action that consumes less energy, resulting in a lower chance of correctly detecting a vehicle. Note that for baseline approaches such as greedy or alternating, the recall value lowers due to increased $T_{down}$, as shown in Fig. \ref{fig:static}(b). For example, the greedy policy will result in the best possible average recall when the cost of transmitting is low. The opposite happens to the greedy policy when the energy cost is high, and cameras experience downtime of nine hours per day.

By comparing the performance of the proposed solutions to the alternating policy, we show that by selecting policy intelligently, it is possible to increase recall while simultaneously lower $T_{down}$ as shown in Fig. \ref{fig:static}(b). In its essence, the alternating policy represents results a system would obtain if actions would be selected randomly. On the other hand, the threshold policy seems reasonable. Still, it does not perform well when the cost of transmission is low and also leads to a relatively high daily downtime when the energy cost of transmission is high.
The difference between the two proposed solutions is that one of them receives additional information in a form of estimated energy. Consequently, while the estimated energy has minimal impact on the achieved recall value, the $T_{down}$ is significantly reduced in comparison when we use \ac{gp} to estimate collected energy.

\subsection{Dynamically Varying the Cost of Transmission}

\begin{figure}
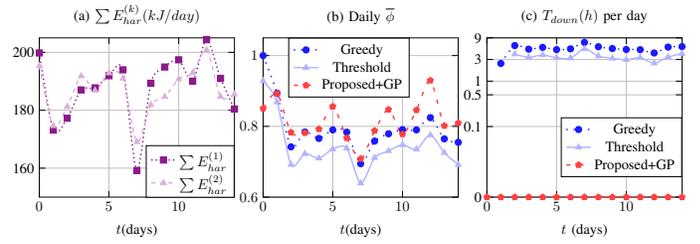

	\centering
	\includestandalone[width=0.5\textwidth]{tikz_figures/daily_performance}
	\caption{The impact of daily collected energy on the performance of the system over a number of days.}
	\label{fig:daily_performance}
	\vspace{-15pt}
\end{figure}

Next, we observe recall and daily downtime as the energy cost of transmission randomly varies over time. At every decision epoch the energy cost $E_{tr}$ is randomly sampled from an interval between $25nJ$ and $85nJ$, i.e, $E_{tr} \in [25,85] nJ$. In Fig.~\ref{fig:daily_performance} we show how the recall and $T_{down}$ change over several days for three selected policies: greedy, proposed with added estimated energy using \ac{gp}, and threshold. In Fig.~\ref{fig:daily_performance}(a) shows how much energy each camera collects in a day. On average, camera one collects more energy. However, the difference is minimal. In Fig.~\ref{fig:daily_performance}(b), we show recall value. Our solution is better at compensating days when cameras obtain less energy, e.g., both cameras collect significantly less energy on day seven than on other days. In Fig.~\ref{fig:daily_performance}(c) we  plot downtime. While the proposed solution results in zero downtime, the greedy approach and threshold directly correlate to collected energy, a trend that is especially noticeable on day seven. 

\begin{figure}
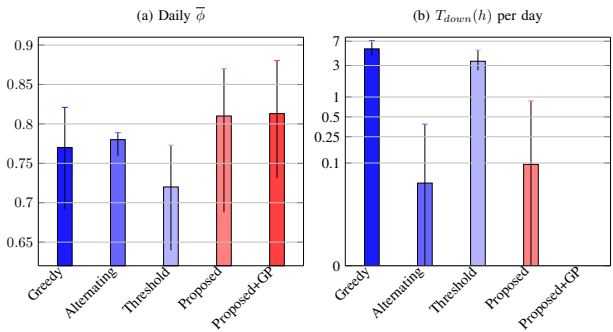

	\centering
	\includestandalone[width=0.45\textwidth]{tikz_figures/bar_plot}
	\caption{Achieved daily average recall and downtime for all approaches when the energy cost of transmission randomly varies over time.}
	\label{fig:bar_plot}
	\vspace{-15pt}
\end{figure}

Fig. \ref{fig:bar_plot} presents the achieved daily values of average recall and downtime for each policy when the cost of transmissions randomly varies. The two proposed solutions acquire the best recall values. However, daily variations are lower when we use \ac{drl}-based approach with estimated energy values.
Interestingly, the threshold policy results in the worst recall values. We plot daily downtime in Fig. \ref{fig:bar_plot}(b). The greedy approach results in the highest downtime per day. Interestingly, the greedy achieves recall similar to the alternating policy. The latter has almost a negligible downtime of only three minutes per day.
Similarly, we could state for our proposed approach as it results in a downtime of around 6 minutes per day. Nevertheless, when we add additional information regarding the incoming energy, the proposed solution will result in no downtime whatsoever. In practice, the system always needs some energy to act in case of an emergency. Therefore, the proposed solution in combination with \ac{gp} is the most beneficial as it will ensure that system will always have some energy left to operate.

%% file: tables/simulation_parameters.tex
\begin{tabular}{llllll}
		\toprule
		\begin{tabular}[c]{@{}c@{}} Parameter\end{tabular} & 
		\begin{tabular}[c]{@{}c@{}} Value \end{tabular} &
		\begin{tabular}[c]{@{}c@{}} Parameter\end{tabular} & 
		\begin{tabular}[c]{@{}c@{}} Value \end{tabular} &
		\begin{tabular}[c]{@{}c@{}} Parameter\end{tabular} & 
		\begin{tabular}[c]{@{}c@{}} Value \end{tabular} \\
		\arrayrulecolor{black}\hline 
		\midrule 
		
		\begin{tabular}[c]{@{}c@{}}  $K$ \end{tabular} & 
		\begin{tabular}[c]{@{}c@{}}  2 \end{tabular} &
		
		\begin{tabular}[c]{@{}c@{}}  $\lambda$ \end{tabular} & 
		\begin{tabular}[c]{@{}c@{}}  $10$\end{tabular} &
		
		\begin{tabular}[c]{@{}c@{}}  $\tau$ \end{tabular} & 
		\begin{tabular}[c]{@{}c@{}}  $1s$\end{tabular} \\
		
			\begin{tabular}[c]{@{}c@{}}  $U_{eh}^{(k)}(t)$ \end{tabular} & 
		\begin{tabular}[c]{@{}c@{}}  $6.1V$ \end{tabular} &
		
		\begin{tabular}[c]{@{}c@{}}  $E_{max}$ \end{tabular} & 
		\begin{tabular}[c]{@{}c@{}}  $185 kJ$ \end{tabular} &
		
		\begin{tabular}[c]{@{}c@{}}  $E_{op}^{(k)}$\end{tabular} & 
		\begin{tabular}[c]{@{}c@{}}  $139mJ$ \end{tabular} \\
		
		\begin{tabular}[c]{@{}c@{}}  $\eta_{eh}$ \end{tabular} & 
		\begin{tabular}[c]{@{}c@{}}  0.8 \end{tabular} &
		
		\begin{tabular}[c]{@{}c@{}} $E_{det}$  \end{tabular} & 
		\begin{tabular}[c]{@{}c@{}}  $57.48mJ$\end{tabular} &
		
		\begin{tabular}[c]{@{}c@{}}  $\tau$ \end{tabular} & 
		\begin{tabular}[c]{@{}c@{}}  $1s$ \end{tabular} \\

		\bottomrule   
\end{tabular}

%% file: tikz_figures/static_cost_of_transmitting.tex
 \begin{tikzpicture}
\begin{groupplot}[group style={group name=my_plots, group size=2 by 1,horizontal sep=1.25cm}, width=0.4\textwidth]]

 \nextgroupplot[
title = (a) $\overline{\phi}$,
xlabel=$E_{tr}(nJ/b)$,
xmin= 0,xmax= 14,
grid=both,
minor grid style={gray!35},
xtick ={25.0, 35.0, 45.0, 55.0, 65.0, 75.0, 85.0},
xticklabels={25, 35, 45, 55, 65, 75, 85},
legend style={at={(1.15,1)}},
xmin=22,xmax=87,
ymin=0.575,ymax=1.15,
legend columns=2, 
every tick/.style={
	black,
	semithick,
}]

\addplot[style=loosely dotted, mark size=2pt, line width=1pt, color=blue!90, mark=otimes*, smooth, mark options={solid}]
    table[x=energy,y=acc,col sep=comma]{csv_data/greedy.csv};

\addplot[style= dashed, mark size=2pt, line width=1pt, color=red!50, mark=diamond*, smooth, mark options={solid}]
	table[x=energy,y=acc,col sep=comma]{csv_data/4_states_deep_learning.csv};

\addplot[style=dotted,mark size=2pt, line width=1pt, color=blue!60, mark=square*, smooth, mark options={solid}]
	table[x=energy,y=acc,col sep=comma]{csv_data/random.csv};

\addplot[style=loosely dashed, mark size=2pt, line width=1pt, color=red!75, mark=pentagon*, smooth, mark options={solid}]
    table[x=energy,y=acc,col sep=comma]{csv_data/deep_learning.csv};
	
\addplot[style=solid,mark size=2pt, line width=1pt, color=blue!30, mark=triangle*, smooth, mark options={solid}]
	table[x=energy,y=acc,col sep=comma]{csv_data/threshold.csv};	

\addlegendentry{Greedy};
\addlegendentry{Proposed};
\addlegendentry{Alternating};
\addlegendentry{Proposed+GP};
\addlegendentry{Threshold};

 \nextgroupplot[
	ymode=log,
	xlabel=$E_{tr}(nJ/b)$, 
	title = (b) $T_{down}(h/day)$,
	grid=both,
	major grid style={gray!35},
	ytick ={10, 360, 900, 1800, 3600, 10800, 32400},
    yticklabels={0, 0.1, 0.25, 0.5, 1, 3, 9},
	xtick ={25.0, 35.0, 45.0, 55.0, 65.0, 75.0, 85.0},
    xticklabels={25, 35, 45, 55, 65, 75, 85},
	legend style={at={(1.2,0.55)}},
	xmin=20,xmax=90,
	scaled y ticks = false,
	ymin=10, ymax=45000,
	every tick/.style={
	        black,
	        semithick,
	}]

\addplot[style=loosely dotted, mark size=2pt, line width=1pt, color=blue!90, mark=otimes*, smooth, mark options={solid}]
    table[x=energy,y=down_time,col sep=comma]{csv_data/greedy.csv};

\addplot[style=dotted,mark size=2pt, line width=1pt, color=blue!60, mark=square*, smooth, mark options={solid}]
	table[x=energy,y=down_time,col sep=comma]{csv_data/random.csv};

\addplot[mark size=2pt, line width=1pt, color=blue!30, mark=triangle*, mark options={solid}]
	table[x=energy,y=down_time,col sep=comma]{csv_data/threshold.csv};	

\addplot[style= dashed, mark size=2pt, line width=1pt, color=red!50, mark=diamond*, smooth, mark options={solid}]
	table[x=energy,y=down_time,col sep=comma]{csv_data/4_states_deep_learning.csv};

\addplot[style=loosely dashed, mark size=2pt, line width=1pt, color=red!75, mark=pentagon*, smooth, mark options={solid}]
    table[x=energy,y=down_time,col sep=comma]{csv_data/deep_learning.csv};

\addlegendentry{Greedy};
\addlegendentry{Alternating};
\addlegendentry{Threshold};
\addlegendentry{Proposed};
\addlegendentry{Proposed+GP};

 
 \end{groupplot}  
 \end{tikzpicture}

%% file: tikz_figures/daily_performance.tex

\begin{tikzpicture}
\begin{groupplot}[group style={group name=my_plots, group size=3 by 1,horizontal sep= 0.7cm}, width=0.35\textwidth]]

\nextgroupplot[
title = (a) $\sum E_{har}^{(k)}(kJ/day)$,
xlabel=$t$(days),
xmin= 0,xmax= 14,
ytick ={160000, 180000, 200000},
yticklabels={160, 180, 200},
scaled y ticks = false,
legend style={at={(1,0.325)}},
ymin= 150000, ymax= 205000,
grid=both,
minor grid style={gray!35},
every tick/.style={
	black,
	semithick,
}]


\addplot[color=violet!90,, dotted, line width=1pt, mark=square*, smooth, mark options={solid}] 
	table[x=days,y=available_energy,col sep=comma]{csv_data/daily_cam1.csv};
	
\addplot[color=violet!30,, dashed, line width=1pt, mark=triangle*, smooth, mark options={solid}] 
table[x=days, y=available_energy, col sep=comma]{csv_data/daily_cam2.csv};

\addlegendentry{$\sum E_{har}^{(1)}$};
\addlegendentry{$\sum E_{har}^{(2)}$};




\nextgroupplot[
title = (b) Daily $\overline{\phi}$,
xlabel=$t$(days),
xmin= 0,xmax= 14,
grid=both,
minor grid style={gray!35},
legend style={at={(0.75,1)}},
ymin= 0.6,ymax= 1.05,
every tick/.style={
	black,
	semithick,
}]



\addplot[style=loosely dotted, mark size=2pt, line width=1pt, color=blue!90, mark=otimes*, smooth, mark options={solid}]
table[x=days, y=accuracy_mean, col sep=comma]{csv_data/daily_greedy.csv};
\addplot[style=solid,mark size=2pt, line width=1pt, color=blue!30, mark=triangle*, smooth, mark options={solid}]
table[x=days, y=accuracy_mean, col sep=comma]{csv_data/daily_threshold.csv};
\addplot[style=loosely dashed, mark size=2pt, line width=1pt, color=red!75, mark=pentagon*, smooth, mark options={solid}]
table[x=days, y=accuracy_mean, col sep=comma]{csv_data/daily_deep_learning.csv};

\addlegendentry{Greedy};
\addlegendentry{Threshold};
\addlegendentry{Proposed+GP};
\nextgroupplot[
title = (c) $T_{down}(h)$ per day,
xlabel=$t$ (days),
ymode=log,
xmin= 0,xmax= 14,
scaled y ticks = false,
ytick ={10, 360, 1800, 3600, 10800, 32400},
yticklabels={0, 0.1, 0.5, 1, 3, 9},
legend style={at={(1,0.5)}},
grid=both,
minor grid style={gray!35},
ymin= 10, ymax= 32400,
every tick/.style={
	black,
	semithick,
}]



\addplot[style=loosely dotted, mark size=2pt, line width=1pt, color=blue!90, mark=otimes*, smooth, mark options={solid}]
table[x=days, y=both_cameras_down, col sep=comma]{csv_data/daily_greedy.csv};
\addplot[style=solid,mark size=2pt, line width=1pt, color=blue!30, mark=triangle*, smooth, mark options={solid}]
table[x=days, y=both_cameras_down, col sep=comma]{csv_data/daily_threshold.csv};
\addplot[style=loosely dashed, mark size=2pt, line width=1pt, color=red!75, mark=pentagon*, smooth, mark options={solid}]
table[x=days, y=both_cameras_down, col sep=comma]{csv_data/daily_deep_learning.csv};

\addlegendentry{Greedy};
\addlegendentry{Threshold};
\addlegendentry{Proposed+GP};


\end{groupplot}
\end{tikzpicture}

%% file: tikz_figures/bar_plot.tex
\pgfplotstableread{
1 0.77 0.051 0.079 19283 6438 4244
}\datasetGREEDY

\pgfplotstableread{
1.5 0.78 0.009 0.021 179 1218 169
}\datasetALTER

\pgfplotstableread{
2 0.72 0.053 0.081 12535 6015 3573
}\datasetTHOLD

\pgfplotstableread{
2.5 0.81 0.06 0.123 344 2809 334
}\datasetLEARNBASIC

\pgfplotstableread{
3 0.813 0.0675 0.082 0 0 0
}\datasetLEARN

\begin{tikzpicture}
\begin{groupplot}[group style={group name=my_plots, group size=2 by 1,horizontal sep=1cm}, width=0.35\textwidth]]

\nextgroupplot[
        scale only axis,
        clip=false,
        separate axis lines,
        axis on top,
        ymajorgrids,
	    minor grid style={gray!35},
        xmin=0.75,
        xmax=3.25,
        xtick={1,1.5,2,2.5,3},
        x tick style={draw=none},
        xticklabels={Greedy, Alternating, Threshold, Proposed, Proposed+GP},
        xticklabel style = {rotate=45,anchor=east},
        ymin=0.62,
        ymax=0.91,
        title = (a) Daily $\overline{\phi}$,
        every axis plot/.append style={
          ybar,
          bar width=10,
          bar shift=0pt,
          fill
        }]
      
    
      \addplot[blue!90, draw=black, error bars/.cd,y dir=both,y explicit] 
      table[x index=0,y index=1, y error plus index=2,y error minus index=3] \datasetGREEDY;

      \addplot[blue!60, draw=black, ybar, error bars/.cd,y dir=both,y explicit] 
      table[x index=0,y index=1, y error plus index=2,y error minus index=3] \datasetALTER;
      
      \addplot[blue!30, draw=black, ybar, error bars/.cd,y dir=both,y explicit]
      table[x index=0,y index=1, y error plus index=2,y error minus index=3] \datasetTHOLD;
      
      \addplot[red!50, draw=black, ybar, error bars/.cd,y dir=both,y explicit] 
      table[x index=0,y index=1, y error plus index=2,y error minus index=3] \datasetLEARNBASIC;
      
      \addplot[red!75, draw=black, ybar, error bars/.cd,y dir=both,y explicit]
      table[x index=0,y index=1, y error plus index=2,y error minus index=3] \datasetLEARN;

\nextgroupplot[
        ymode=log,
        scale only axis,
        clip=true,
        separate axis lines,
        axis on top,
        ymajorgrids,
	    minor grid style={gray!35},
        xmin=0.75,
        xmax=3.25,
        xtick={1,1.5,2,2.5,3},
        x tick style={draw=none},
        xticklabels={Greedy, Alternating, Threshold, Proposed, Proposed+GP},
        xticklabel style = {rotate=45,anchor=east},
        ytick ={10, 360, 900, 1800, 3600, 10800, 25200},
        yticklabels={0, 0.1, 0.25, 0.5, 1, 3, 7},
        ymin=10, ymax=29000,
        title = (b) $T_{down}(h)$ per day,
        every axis plot/.append style={
          ybar,
          bar width=10,
          bar shift=0pt,
          fill
        }]
      
      \addplot[blue!90, draw=black, ybar, error bars/.cd,y dir=both,y explicit]
      table[x index=0,y index=4, y error plus index=5,y error minus index=6] \datasetGREEDY;
      \addplot[blue!60, draw=black, ybar, error bars/.cd,y dir=both,y explicit]
      table[x index=0,y index=4, y error plus index=5,y error minus index=6] \datasetALTER;
      \addplot[blue!30, draw=black, ybar, error bars/.cd,y dir=both,y explicit]
      table[x index=0,y index=4, y error plus index=5,y error minus index=6] \datasetTHOLD;
      \addplot[red!50, draw=black, ybar, error bars/.cd,y dir=both,y explicit] 
      table[x index=0,y index=4, y error plus index=5,y error minus index=6] \datasetLEARNBASIC;
      \addplot[red!75, draw=black, ybar, error bars/.cd,y dir=both,y explicit]
      table[x index=0,y index=4, y error plus index=5,y error minus index=6] \datasetLEARN;

  \end{groupplot}  
  \end{tikzpicture}

%% file: src/sections/7_conclusion.tex

In this paper, we studied how can sensor network take advantage of correlation to improve the energy efficiency of \ac{eh}-powered devices without impacting performance. We showed the benefits of utilising correlation in a scenario of multiple cameras powered only by \ac{eh}, observing the same intersection with a goal to count the number of vehicles accurately.We proposed a \ac{drl}-based approach to select cameras' actions to improve system performance we measure through recall while preventing outages. Our results show that the proposed approach can be up to $15\%$ more accurate with no outages. With millions of \ac{eh} powered devices in future networks, the use of correlated information will become increasingly more important to balance the energy performance of such devices while maximising their performance.


In our future work, we will examine the scalability of our solution. In theory, the more cameras the system has, the more advantageous should it be to leverage correlation. However, as the number of cameras increases, so does the number of available actions; thus, a solution capable of dealing with a high number of actions is required. We will also investigate the delay. The camera needs more time to process the image than a cloudlet server. In contrast, it takes more time to transmit raw image than the result of the data analytics task. By considering the delay, we will add another dimension to our work. Therefore, we expect to observe interesting results when considering three tradeoffs: energy, accuracy, and delay.